\documentclass[twocolumn]{aastex61}
\pdfoutput=1 
\usepackage{amsmath,amstext}
\usepackage[T1]{fontenc}
\usepackage{apjfonts} 
\usepackage[figure,figure*]{hypcap}

\shorttitle{Instabilities and Clump Sizes}
\shortauthors{Fisher et al.}

\begin{document}

\title{Connecting Clump Sizes in Turbulent Disk Galaxies to Instability Theory}

\author{David~B.~Fisher}
\affiliation{Centre for Astrophysics and Supercomputing, Swinburne
  University of Technology, P.O. Box 218, Hawthorn, VIC 3122,
   Australia}
   
\author{Karl Glazebrook}
\affiliation{Centre for Astrophysics and Supercomputing, Swinburne
  University of Technology, P.O. Box 218, Hawthorn, VIC 3122,
   Australia}
\affiliation{ARC Centre of
   Excellence for All-sky Astrophysics (CAASTRO)} 
   
\author{Roberto G. Abraham}
\affiliation{Department of Astronomy \& Astrophysics, University of Toronto, 50 St. George St., Toronto, ON M5S 3H8, Canada}
  
\author{Ivana Damjanov}
\affiliation{Harvard-Smithsonian Center
 for Astrophysics, 60 Garden St., Cambridge, MA 02138,
  USA}
  
\author{Heidi White}
\affiliation{Department of Astronomy \& Astrophysics, University of Toronto, 50 St. George St., Toronto, ON M5S 3H8, Canada}

\author{Danail Obreschkow}
\affiliation{International Centre for Radio Astronomy Research (ICRAR), M468, University of Western Australia, 35 Stirling Hwy, Crawley, WA
6009, Australia}

\author{Robert Basset}
\affiliation{International Centre for Radio Astronomy Research (ICRAR), M468, University of Western Australia, 35 Stirling Hwy, Crawley, WA
6009, Australia}

\author{Georgios Bekiaris}
\affiliation{Centre for Astrophysics and Supercomputing, Swinburne
  University of Technology, P.O. Box 218, Hawthorn, VIC 3122,
   Australia}
   
\author{Emily Wisnioski}   
\affiliation{Max-Planck-Institut f\"ur extraterrestrische Physik, Postfach
 1312, Giessenbachstr., D-85741 Garching, Germany}

\author{Andy Green}
\affiliation{Australian Astronomical Observatory, P.O. Box 970,
North Ryde, NSW 1670, Australia}

\author{Alberto~D.~Bolatto}
\affiliation{Laboratory of Millimeter Astronomy, University of Maryland, College
Park, MD 29742}

\begin{abstract}
  In this letter we study the mean sizes of H$\alpha$ clumps in
  turbulent disk galaxies relative to kinematics, gas fractions, and
  Toomre $Q$.  We use $\sim100$~pc resolution HST images, IFU
  kinematics, and gas fractions of a sample of rare, nearby turbulent
  disks with properties closely matched to $z\sim1.5-2$ main-sequence
  galaxies (the DYNAMO sample).  We find linear correlations of
  normalized mean clump sizes with both the gas fraction and the
  velocity dispersion-to-rotation velocity ratio of the host
  galaxy. We show that these correlations are consistent with
  predictions derived from a model of instabilities in a
  self-gravitating disk (the so-called ``violent disk instability
  model'').  We also observe, using a two-fluid model for $Q$, a
  correlation between the size of clumps and self-gravity driven unstable regions.
  These results are most consistent with the hypothesis that massive star forming clumps in turbulent disks are the result of instabilities in self-gravitating gas-rich disks, and therefore provide a direct connection between resolved
  clump sizes and this {\em in situ} mechanism.
\end{abstract}


\section{Introduction}
Galaxies in the distant past ($z>1.5$) were dominated by massive
star-forming ``clumps'', where star formation rates in clumps are
$\sim$500 times higher than star-forming regions in the Milky Way
\citep{elmegreen2005,genzel2011,swinbank2012,wisnioski2012,guo2012}. The
``clumpiness'' of galaxies is linked to star formation rate density of
the Universe \citep{shibuya2015arxiv}. 
However, due to current limits on
resolution and sensitivity links between observed clump properties and
theories of clump formation remain qualitative in nature.

 Some authors propose
that the large star forming regions in clumpy, turbulent galaxies are
the result of instabilities which originate via self-gravity
\citep{dekel2009clumps,genzel2011} similar to spiral galaxies in the
local Universe\citep{kennicutt1989,elmegreen1991,wada2002}. However, drastic differences between nearby spirals
and turbulent disks \citep{forsterschreiber2009,wisnioski2015} could
possibly point to different mechanisms driving star
formation. Alternatively, instabilities leading to clumps could be
driven by interactions with other galaxies in which gas rich
interactions maintain a disk\citep{robertson2006}. 

The most popular theory for clump formation is {\em self-gravitating
  disk instability model}. In this model, above a certain size
rotation stabilizes against gravity driven fragmentation. The critical
size is therefore the largest size structure that will form in an
unstable disk.  For circular orbits this {\em rotation size scale}
\citep{toomre1964,binneyandtremaine} can be expressed as
\begin{equation}
R_{rot} \propto \frac{ G \Sigma}{\Omega^2},
\label{eq:rotscale}
\end{equation}
where $\Omega$ is the angular rotation velocity of the disk, $\Sigma$
is the surface density of gas, and G is the gravitational
constant. If self-gravity driven
instabilities form clumps, then the size of the clumps must
obey the relationship in Equation~1.

\begin{figure*}
\begin{center}
\includegraphics[width=0.9\textwidth]{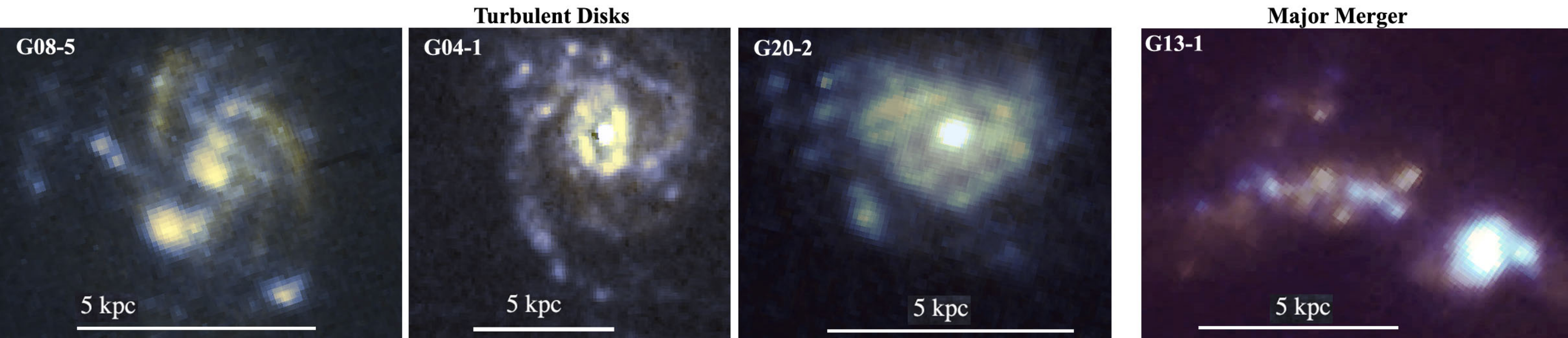} 
\end{center}
\caption{ Two color HST images of DYNAMO galaxies are shown. Blue
  represents H$\alpha$ and yellow represents 600~nm continuum. The
  H$\alpha$ clumps in DYNAMO
  galaxies have been shown to be very similar in properties to those
  in high redshift galaxies. The left three galaxies are identified as clumpy disks, for comparison we also show G13-1 a galaxy classified as an ongoing merger in the DYNAMO-HST sample. \label{fig:maps}  }
\end{figure*} 
\begin{deluxetable*}{lcccccccc}
\tablewidth{0pt} \tablecaption{Properties of DYNAMO-{\em HST} Sample}
\tablehead{ \colhead{Galaxy} & \colhead{z} & \colhead{SFR} & \colhead{M$_{star}$} &\colhead{Type} &
  R$_{1/2}$(disk)\tablenotemark{a} & <R$_{clump}>\tablenotemark{b}$ & $\sigma/V$ & f$_{gas}$\tablenotemark{c} \\
\colhead{ }& \colhead{ } &\colhead{$M_{\odot}$ yr$^{-1}$} & \colhead{log($M_{\odot}$)} &\colhead{ } & \colhead{ kpc } & \colhead
{kpc} & \colhead{ } & \colhead { }
}
\startdata
G04-1 & 0.12981 & 21.32 & 10.81 & disk & 2.75 & 0.46 $\pm$ 0.26 & 0.19 $\pm$ 0.09 & 0.33 $\pm$ 0.04 \\
G20-2 & 0.14113 & 18.24 & 10.33 & disk & 2.1 & 0.66 $\pm$ 0.29 & 0.49 $\pm$ 0.07 & 0.21 $\pm$ 0.05 \\
D13-5 & 0.0753 & 17.48 & 10.73 & disk & 2.04 & 0.41 $\pm$ 0.15 & 0.24 $\pm$ 0.02 & 0.36 $\pm$ 0.02 \\
G08-5 & 0.13217 & 10.04 & 10.24 & disk & 1.84 & 0.37 $\pm$ 0.07 & 0.26 $\pm$ 0.07 & 0.3 $\pm$ 0.05 \\
D15-3 & 0.06712 & 8.29 & 10.73 & disk & 2.2 & 0.22 $\pm$ 0.09 & 0.19 $\pm$ 0.02 & 0.17 $\pm$ 0.02 \\
G14-1 & 0.13233 & 6.90 & 10.35 & disk & 1.12 & 0.45 $\pm$ 0.17 & 0.51 $\pm$ 0.07 & 0.77 $\pm$ 0.08 \\
C13-1 & 0.07876 & 5.06 & 10.55 & disk & 4.21 & 0.34 $\pm$ 0.11 & 0.13 $\pm$ 0.04 & 0.06 $\pm$ 0.02 \\
A04-3 & 0.06907 & 2.42 & 10.63 & disk & 3.58 & 0.17 $\pm$ 0.10 & 0.05 $\pm$ 0.03 & .01 $\pm$ 0.01 \\
H10-2 & 0.14907 & 15.49 & 9.98 & merger & 2.55 & 0.82 $\pm$ 0.21 & 0.95 $\pm$ 0.34 & $<$0.67   \\
G13-1 & 0.13876 & 12.71 & 10.05 & merger & 2.59 & 0.55 $\pm$ 0.13 & 0.68 $\pm$ 0.03 & $<$0.13   \\
\enddata
\tablenotetext{a}{$R_{disk}$ is measured as twice $R_{1/2}$ for H$\alpha$ light.}
\tablenotetext{b}{$R_{clump}$ represents an average for each galaxy.}
\tablenotetext{c}{Galaxy A04-3 does not have a CO(1-0)
  measurement, for this target we assume that M$_{gas}=10^9 \times
  $SFR, where SFR is calculated from the extinction corrected
  H$\alpha$ luminosity using \cite{hao2011} calibration.}
\end{deluxetable*}



In this letter we use the
DYNAMO sample \citep{green2010} ({\bf DY}namics of {\bf N}ewly-{\bf
  A}ssembled {\bf M}assive {\bf O}bjects). DYNAMO disks are very
similar to main-sequence disks at $z\sim 1.5-2$, yet DYNAMO galaxies are at $z\approx0.1$.  DYNAMO disk galaxies
have high star formation rates, gas fractions ($f_{gas}\sim 20-40$\%)
\citep{fisher2014} and H$\alpha$ velocity dispersions
(30-80~km~s$^{-1}$) \citep{green2010,bassett2014}, similar to those of
$z\sim 1.5$ turbulent disks. In recent work\citep{fisher2016tmp} we
have shown that 8 of the 10 DYNAMO-HST galaxies (the sample analyzed
in this work) pass definitions of clumpy galaxies used in other
surveys \citep[e.g.][]{guo2015}, and show that when viewed at similar
resolution, DYNAMO disks have essentially identical morphologies as
those of $z\sim1.5-2$ galaxies.

Throughout this paper,
we assume a concordance cosmology with \hbox{$H_0$ = 67 km\ $s^{-1}$ Mpc$^{-1}$}, $\Omega_M=0.31$,
and $\Omega_\Lambda=0.69$.


\section{Methods}

Our data set includes galaxies with H$\alpha$ narrowband imaging from
the {\em Hubble Space Telescope} with 0.05-0.2~kpc resolution. These
DYNAMO-HST galaxies were imaged using the HST ramp filters FR716N and
FR782N, which target the H$\alpha$ emission line with a 2\%
bandwidth. In Fig.~\ref{fig:maps} we show examples of clumpy, dynamo
disks from the DYNAMO-HST sample. This data set, measurement of clump properties, and
disk/merger classification is described in detail in
\citep{fisher2016tmp}. ``Disk galaxies'' in this work are those that
show both signs of rotation and have exponentially decaying stellar
surface brightness profiles.


{\bf Clump Sizes:} Clumps were identified as 3$\sigma$ enhancements in
an unsharp mask image created from the H$\alpha$ maps. We did not
place any size restriction on the sizes of clumps. 
Clumps that are co-located with the galaxy center (in the continuum) are
removed from analysis. In the DYNAMO sample we identified 113 clumps
for study in this work.

\begin{figure*}
\begin{center}
\includegraphics[width=0.9\textwidth]{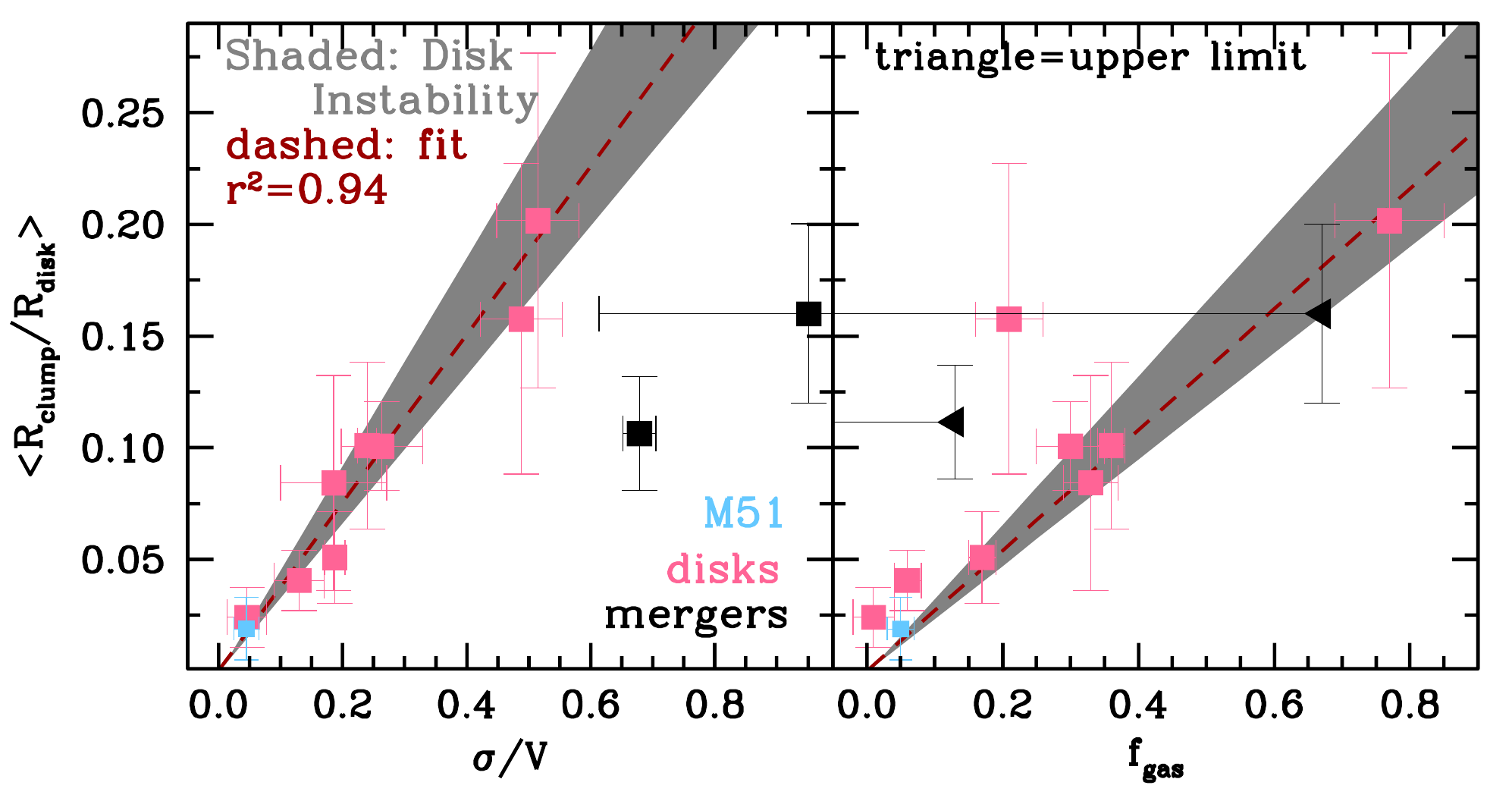} 
\end{center}
\caption{The pink points
  denote DYNAMO disk galaxies. The black points represent the 2 DYNAMO
  galaxies that are consistent with
  mergers\cite{fisher2016tmp}. The light blue point represents
  the same analysis for M51. 
  In each panel the dashed line represents a fit to the disk galaxies
  only. The gray shaded region represents the prediction from the
  violent disk instability model. \label{fig:cors}  }
\end{figure*}

To estimate the sizes of star forming clumps, we fit 
Gaussian functions to the 2-D brightness distributions surrounding
each peak in the H$\alpha$+[NII] map and set $R_{core} = \sqrt{a\times
   b}$, where $a$ \& $b$ are major and minor-axis. 
For our purposes, the
characteristic clump size needs to reflect the full length of the
collapsing region.
We find a very strong correlation between $R_{core}$ and $R_{90}$, where $R_{90}/R_{core} \approx 2$, and $R_{core}$ is
much less affected by noise. We then define the
size scale $R_{clump}=2\times R_{core}$, which for perfect Gaussians
would contain $\sim 95$\% of flux of each clump. 

We define $R_{disk}\equiv 2 R_{1/2}(H\alpha)$ using HST H$\alpha$ maps. Disk surface photometry is determined with elliptical isophote fits using the software described in \cite{fisherdrory2008}.

The median beam size of our maps is 0.088
arcsec, which is 100~pc at $z=0.07$ and 180~pc at $z=0.13$.  We
subtract this beam size in quadrature from the clump sizes base on
H$\alpha$+[NII] maps to yield the final clump size. Values for average
$R_{clump}$ are given in Table~1.

{\bf Kinematics:} We determined kinematic properties, velocity dispersion ($\sigma$),
and rotation velocity ($V$) of the 6 most distant ($z\sim0.14$) DYNAMO
galaxies (D13-5, G04-1, G08-5, G14-1, G20-2 and H10-2) with Gemini
GMOS observations \citep{bassett2014}. The kinematics of the 4 nearer ($z\sim0.07$) DYNAMO
galaxies (A04-3, C13-1, D15-3 and G13-1) were determined with
AAT/WiFes observations \citep{green2014}. The spatial resolution of the WiFes
observations (1.4~arcsec) is lower than that of the GMOS. However,
targets A04-3, C13-1 and D15-3 are roughly a factor of 2 closer
($z\sim0.075$) than most of the galaxies observed with GMOS ($z\sim
0.14$). The physical resolution for both kinematic data
sets is $\sim1-2$~kpc.

Emission line data cubes containing intensity, velocity dispersion and
rotation velocity of each galaxy were then fit with kinematic models
by method of least-squares using the GPU based software {\em
  gbkfit} \citep[see][]{bekiaris2016}. 
We model the rotation velocity, $v_{rot}$, with the
function\citep{boissier2003}
\begin{equation}
v_{rot}(r) = v_{flat} \left [ 1 - exp(-r/r_{flat}) \right].
\end{equation}
Where $v_{flat}$ and $r_{flat}$ are free parameters were
fit to the 2-D velocity profile. The software also fits velocity
dispersion, assuming a flat velocity dispersion profile for the
galaxy.

{\bf Gas Mass:} We follow a similar procedure as in \cite{fisher2014} 
to measure the gas fractions of DYNAMO galaxies. We observed the
CO(1-0) line with Plateau de Bure interferometer in D
configuration. 
Data were
calibrated using standard methods at the IRAM facility, then binned
into 20~km~s$^{-1}$ spectra. 

The CO(1-0) flux was then converted to molecular gas mass ($M_{mol}$)
in the usual fashion, in which $M_{mol} = \alpha_{CO} L_{CO}$, where
$L_{CO}$ is the luminosity of CO(1-0), and $\alpha_{CO}$ is the
CO-to-H$_2$ conversion factor, including a 1.36$\times$ correction for
heavier molecules. We
 adopted the standard value $\alpha_{CO}=4.36$. For more discussion of this conversion factor in DYNAMO galaxies see \cite{fisher2014} and White et al ({\em in prep}).

\section{Results}

\subsection{Instabilities in a self-gravitating disk}
The $Q$ parameter estimates the stability of a gravitating disk
against collapse \citep{toomre1964,safranov1960}. For a gaseous disk
$Q_{gas} = \kappa \sigma / (\pi G \Sigma_{gas})$, where $\kappa$ is
the epicyclic frequency and $\Sigma_{gas}$ is the surface density of
gas. A similar expression for stability exists for $Q_{stars}$. Both
components contribute to the gravitation field and thus both are
important for stability of the disk; these are often combined in the
so-called ``two-fluid'' model such that
$Q^{-1}=Q_{stars}^{-1}+Q_{gas}^{-1}$ \citep{wangsilk1994}. Where
$Q<1$, the disk is unstable.


Under the condition of $Q_{gas}\sim1$, which is consistent with our
data, it is straightforward to derive a direct, observable,
relationship of increasing clump size with both gas fraction and
$\sigma/V$ from Equation~1. We use a derivation that balances the
change in gravitational force of a collapsing element of gas with the
centrifugal force induced by the differential rotation of the gas
disk. We also make the assumption that the appropriate comparison is
the ``most unstable mode'', which is 1/2 the critical size
\citep{binneyandtremaine}.  We find the following relationship,
\begin{equation}
\frac{R_{clump}}{R_{disk}} = a \frac{1}{3} \left( \frac{\sigma}{V}
\right).
\label{eq:rotation_sig}
 \end{equation} 
 The constant $a$ relates $\kappa= a (V/R_{disk})$, and for a typical
 disk varies from $a=1$ to $a=\sqrt2$\citep{binneyandtremaine}.
 Substituting $f_{gas}$ for $\sigma/V$ returns a linear relationship
 between clump size and gas fraction
 \citep{glazebrook2013,genzel2011}. These capture the concept that in
 turbulent disks, the clumps are larger if the gas is more turbulent
 (higher $\sigma/V$) and the disk is more gas rich (higher
 $f_{gas}$). Similar correlations are discussed elsewhere in the
 literature \citep{escala2008,dekel2009clumps,guo2012}. 
 

 In Fig.~\ref{fig:cors} we show that for DYNAMO-HST disk galaxies, there is a very
 strong correlation between average $R_{clump}/R_{disk}$ and
 $\sigma/V$.  The best fit relationships yields $R_{clump}/R_{disk}
 =(0.38\pm0.02)~\sigma/V$, consistent with the range $a=1$ to $a=\sqrt
 2$ in Eq.~\ref{eq:rotation_sig}. The correlation with $f_{gas}$ is less strong. In
 the same figure we show the predictions based on the model of
 instabilities in a self-gravitating disk,
 Eq.~\ref{eq:rotation_sig}. We find that correlations of clump sizes
 in DYNAMO disks with both $\sigma/V$ and $f_{gas}$ are consistent
 with predictions of the so-called ``violent disk instability''
 model. 

 We also find that the  (unnormalized) $R_{clump}$ correlates well with
 $\sigma/V$.  This is important because the relationship shown in Fig.~\ref{fig:cors} may be
 affected by a correlation between $R_{disk}$ and $\sigma/V$. However,
 in our sample the correlation of $R_{clump}$ with $\sigma/V$ is
 roughly as robust, if not slightly stronger ($r^2 = 0.64$) than that
 of $R_{disk}$ with $\sigma/V$ ($r^2=0.59$).  
\begin{figure*}
\begin{center}
\includegraphics[width=0.9\textwidth]{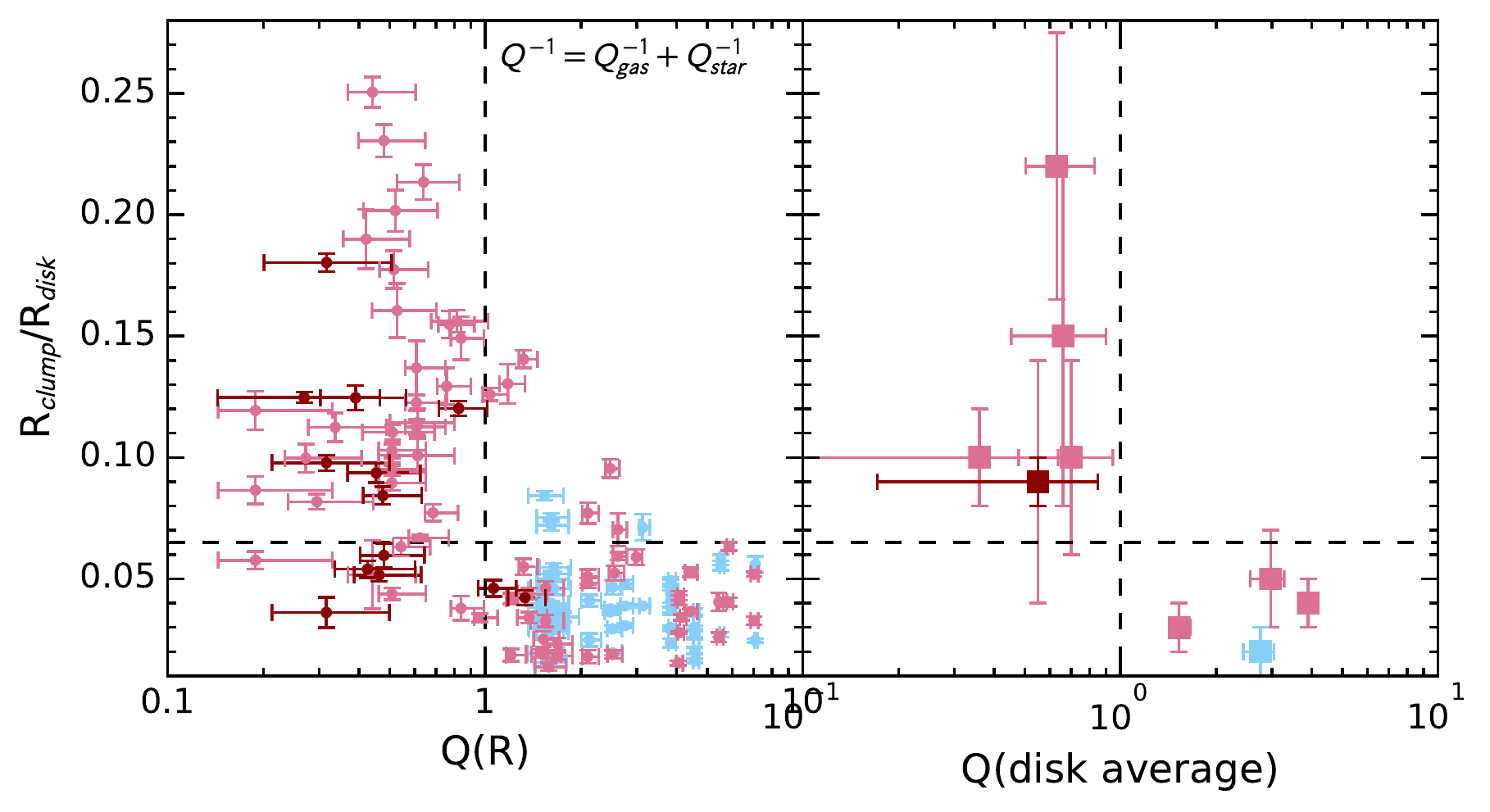} 
\end{center}
\caption{The normalized sizes for all clumps are compared to the value of the
  Toomre stability criteria. The left panel compares individual clump measurements to the azimuthally averaged Q(R). The right panel compares disk averaged values for both $R_{clump}/R_{disk}$ and $Q$. The pink symbols represent clumps in
  the DYNAMO disks, the blue symbols represent those in M51, and the
  dark red symbols represent DYNAMO disk G04-1. The horizontal line represents $R_{clump}/R_{disk}$ value for which 90\% of points with $Q>1$ are smaller than.  \label{fig:qclump}  }
\end{figure*}


 
We note the assumptions for the geometry of the
 model, the method in which Eq.~\ref{eq:rotation_sig} is derived,
 and the measurement of clump sizes may all alter the proportionality
 constant in Eq.~\ref{eq:rotation_sig}. For the most common assumption of a flat
 rotation curve ($\kappa = 2^{1/2}\Omega$) the simple fragmentation
 in a rotating disk and the derivation from the dispersion relation
 yield very similar results, and are both, within error bars,
 consistent with our data. It is, however, possible to construct
 different assumptions that may affect this. Nonetheless, the
 observation of strong, linear correlation between
 $R_{clump}/R_{disk}$ and $\sigma/V$ is itself strongly consistent
 with Eq.~\ref{eq:rotation_sig}independent of the proportionality constant.

To determine the azimuthally
averaged $Q_{gas}(r)$ we directly measure kinematics from the
H$\alpha$ emission line. The gas surface density was determined using
the H$\alpha$ surface density profile and assuming a constant
$M_{gas}/L_{H\alpha}$ ratio across the disk scaled to match our
measurement of molecular gas mass. To ensure that individual $Q(R)$ is not merely a reflection of the clump brightness we compared H$\alpha$ profiles in which clumps are masked to those without masked clumps, and found negligible differences.  To calculate $Q_{star}(r)$ we used
the HST continuum surface brightness profile with a single
mass-to-light ratio determined from SDSS
photometry. \cite{bassett2014} measured stellar kinematics for 2 of
our target galaxies. Based on results from these two targets we
consider the following possibilities: (1) $V_{gas}\approx V_{star}$
and $\sigma_{gas}\approx \sigma_{star}$, (2) $\sigma_{star}\approx
\sigma_{gas}+15$~km~s$^{-1}$, and (3) $\sigma_{star}\approx
1.5\times\sigma_{gas}$). The differences between these three cases are
reflected in the error bars in Fig.~\ref{fig:qclump}.

The self-gravity instability model of clump formation predicts that clumps will form in unstable regions of the disk. In the left panel of we show that in DYNAMO disks large clumps only exists in regions with $Q(R)<1$, and conversely small clumps, $R_{clump}/R_{disk}\leq 3$\%, only reside in stable regions. In Fig.~\ref{fig:qclump} we highlight one
galaxy, G04-1, to show that even within a single disk when $Q$ becomes
stable the clumps become small. Recall that the sub-galactic $Q(R)$ measurement is derived from azimuthally averaged profiles. This $Q(R)$ is intended to reflect the broad region of the disk containing the clump, and not merely the local vicinity of the clump. This is an important distinction as the formation of a clump in a region may lead to that region becoming non-linear in nature \citep{tamburello2015,inoue2016}. 

In Fig.~\ref{fig:qclump} we therefore also show galaxy averaged values for both $Q$ and $R_{clump}/R_{disk}$. The galaxies with low $Q$ have larger clumps. The results
we observe in Fig.~\ref{fig:qclump} builds significantly on compelling early
results\citep{genzel2011,swinbank2012} by controlling for resolution
effects\citep{fisher2016tmp,tamburello2016arxiv}, measuring the full
stability, and quantifiably showing the decrease in clump sizes toward
stable regions. We find strong consistency of clump sizes with
predictions of the model in which self-gravity is the origin of large
clumps.


\subsection{Mergers as drivers of clumps}
{\bf Major Mergers:} In our sample 2 galaxies G13-1 and H10-2 appear most
consistent with a major-merger scenario. Simulations predict that at
high gas fraction major mergers can generate disk like kinematic
profiles \citep{robertson2006} and  morpho-kinematic studies
suggest that a significant fraction of $z>1$ star forming galaxies fit
this picture \citep[recently][]{Rodrigues2017}. 

\begin{figure}
\begin{center}
\includegraphics[width=0.5\textwidth]{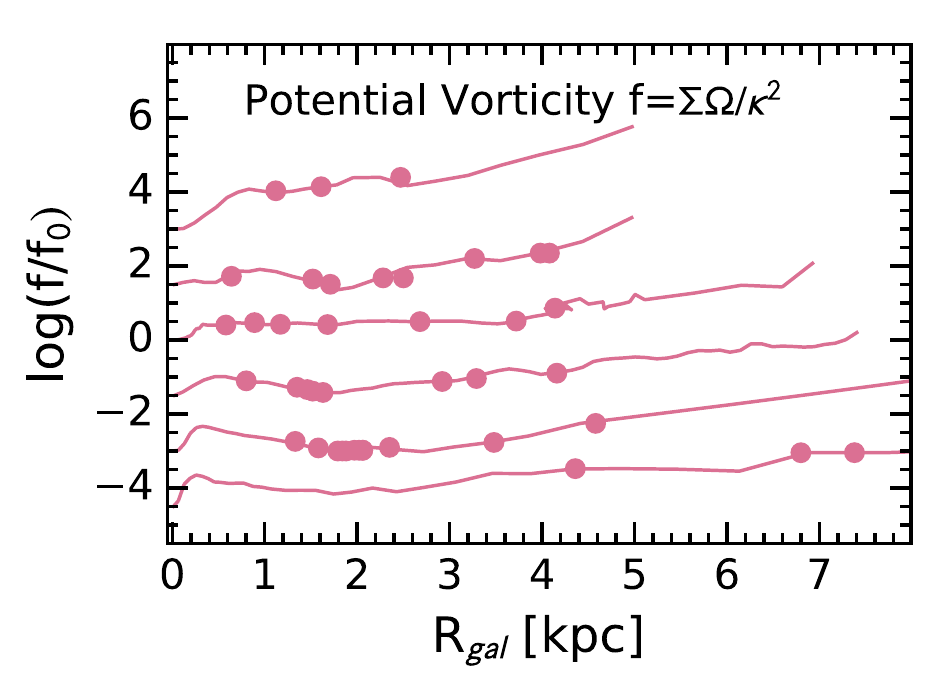} 
\end{center}

\caption{The profiles of potential vorticity each DYNAMO
  of the 6 disks that contains large clumps ($R_{clump}/R_{disk}>7$\%)
  are shown as pink lines. (Note 2 disks A04-3 and C13-1 do not have
  large clumps, and 2 galaxies are mergers.)  The locations of clumps
  are indicated as filled circles. The value of $f_0$ is chosen
  differently for each galaxy to offset the curves, as the gradient is
  all that is important. Theory predicts that non-axisymmetric
  instabilities occur in locations where $f$ is a decreasing function
  of radius. No such preference is seen in our data. \label{fig:fvort}   }
\end{figure}

 If the primary formation mechanism of massive clumps
 were the compression of gas in extremely gas rich major-mergers there is no
 clear reason for the relationships in Fig.~\ref{fig:cors} to hold. Furthermore,
 the effect of mergers on kinematics, even in gas rich mergers, is to
 drive up $\sigma$ significantly. $Q$ then increases, which is
 inconsistent with our result in Fig.~\ref{fig:qclump}. For the same average relative
 clump size DYNAMO merging galaxies (H10-2 and G13-1) are more
 dispersion dominated than the turbulent disks. Though the assumptions
 that are made for determining $Q_{gas}$ do not apply to merging
 galaxies, if one applies this formula to these two merging
 galaxies we find $Q_{gas}\sim$ 6 and
 $Q_{gas}\sim 15$. This is
 consistent with the expectation that those clumpy galaxies that are
 independently identified as merging systems, not disks, are also
 those systems that deviate from the predictions of the self-gravitating disk
 instability model.

{\bf Minor Mergers:}  Assuming that the radial decay of a clump resulting from a minor merger heats the disk as it loses energy, in principle one can derive a similar linear relationship between $R_{clump}/R_{disk}$ and $\sigma/V$. However, we note that many assumptions required in this derivation are significantly uncertain. Though beyond the scope of this work, a comparison of simulation results would be helpful. 
 
A clump with energy, $E_{clump}=1/2 \Sigma_{clump} R_{clump}^2 V_{disk}^2$, will loose energy as it radially decays. This energy is translated to an increase in disk velocity dispersion, $\sigma$.  Using the average position of clumps in clumpy DYNAMO disks (2.7~kpc), the average size of disks in the same galaxies (5~kpc), and assuming $V\propto R^{1/2}$, one derives $\Sigma_{disk} R_{disk}^2 \sigma^2 = 0.51 \Sigma_{clump} R_{clump}^2 V_{disk}^2$. The assumption that $V\propto R^{1/2}$ neglects dark-matter and is thus questionable, however is necessary to derive $R_{clump}/R_{disk}\propto \sigma/V$. The quantity $\Sigma_{disk}/\Sigma_{clump}$ in gas mass or baryonic mass is not known. For DYNAMO H$\alpha$ measurements we find  $\Sigma_{disk}/\Sigma_{clump}\approx 0.4$. Note this assumes that the conversion of H$\alpha$ flux to gas mass is the same for both the clump and the disk, which is also uncertain. This returns the relationship $\frac{R_{clump}}{R_{disk}} \approx 0.9 \frac{\sigma}{V}$, which is much steeper than our result in  Fig.~\ref{fig:cors}. To match our observations $\Sigma_{disk}/\Sigma_{clump}\approx 0.05-0.1$ would be necessary. 
 
This derivation for minor-merger scenarios relies on multiple assumptions that may not apply to real galaxies. First, DYNAMO disks contain $\sim10-15$ clumps per galaxy. \cite{fisher2016tmp} shows that the larger number of clumps compared to $z\sim2$ observations is likely a resolution effect. It seems unlikely that each clump represents a minor merger within the past $\sim$1~Gyr. Moreover, the local gravity and shear are very likely to alter the structure of a clump initiating from a minor merger. An alternate possibility is that clumps resulting from minor mergers reshape to match the stability conditions derived in Eq.~\ref{eq:rotation_sig}. Ultimately a measurement of clump properties in Figs.~\ref{fig:cors} and \ref{fig:qclump} derived from minor mergers in simulations with realistic conditions would be useful. 


To zeroth order, if all clumps result from minor-mergers there should be no dependence of $R_{clump}/R_{disk}$ on galactocentric radius. Conversely, in the disk-instability model $\sigma/V$ should decrease with radius, until $V(R)$ flattens, and therefore $R_{clump}/R_{disk}$ would decrease with galactic radius. In DYNAMO we find that clumps located in the central 20\% of the disk have a median clump size of $R_{clump}/R_{disk}\approx 0.25$, where as the median of the rest of the disk is $R_{clump}/R_{disk}\approx 0.17$. 
 
 
\subsection{Non-axisymmetric instabilities}
 Yet another possible mechanism driving clump formation are
 non-axisymmetric instabilities\citep{lovelacehohlfeld2013}.  These
 instabilities do not require low $Q$, and thus it seems unlikely that
 we would observe such regularity in Fig.~\ref{fig:cors} in this case.  The main
 criterion for non-axisymmetric instabilities is that the {\em
   potential vorticity} be a decreasing function of radius. For
 rotating disk galaxies potential vorticity, $f$, is defined as
\begin{equation}
f \equiv \frac{\Sigma \Omega}{\kappa^2}.
\end{equation}
In Fig.~\ref{fig:fvort} we plot profiles of $f$ for all DYNAMO
galaxies that contain large clumps ($R_{clump}/R_{disk}>7$\%). 
If non-axisymmetric instabilities were driving
clump formation, then we would expect clumps to be associated with
decreasing $\nabla$f. Though some clumps are in decreasing sections of
the profile, there is clearly no preferential relationship between
clump locations and $\nabla$f, suggesting non-axisymmetric
instabilities are not likely driving clump formation.


\section{Discussion}
In this letter we report the observation of positive, linear
correlations between the average, normalized size of clumps
($R_{clump}/R_{disk}$) in turbulent disks with both the ratio of
velocity dispersion-to-rotation speed ($\sigma/V$) of the disk and the
gas fraction ($f_{gas}$) of the disk. These correlations in our data
(Fig.~\ref{fig:cors}) are consistent with predicted relationships that
are derived from model in which clumps form via {\em in situ}
instabilities in a self-gravitating disk (Eq.~\ref{eq:rotation_sig}).
We also find a correlation between the size of clumps and 
Toomre $Q$ value such that large clumps are restricted to regions of
the disk with $Q<1$ (Fig.~\ref{fig:qclump}). 

 
Minor-mergers could lead to a linear relationship between $R_{clump}/R_{disk}$ and $\sigma/V$; however, at present it is not clear that the resulting correlation, nor assumptions necessary match real galaxies. 
Clump formation as
 a simple response to instabilities in 
 self-gravitating disks is the most consistent model with our data.


DYNAMO disks are very similar to
$z\approx1.5-2.0$ main-sequence galaxies
\cite{green2014,fisher2014,fisher2016tmp}.
A common picture of
galaxy evolution, is that the tightness of the star forming main
sequence at a particular redshift intervals indicates that star
formation in these galaxies is driven by a slow extended mechanism
\citep[e.g.][]{daddi2007,dekel2009clumps,wuyts2011}.  Our results
therefore demonstrate a rigorous, quantifiable and direct connection between the
clumpy mode of star formation that dominates $z\sim1-3$ main-sequence
galaxies and a such a mechanism that occurs naturally in gas rich,
turbulent disks. \\ 

\acknowledgments DBF acknolwedges support from Australian Research
Council (ARC) Discovery Program (DP) grant DP130101460.  Support for
this project is provided in part by the Victorian Department of State
Development, Business and Innovation through the Victorian
International Research Scholarship (VIRS).  ADB acknowledges partial
support form AST1412419. Based on observations made with the NASA/ESA
Hubble Space Telescope, obtained from the data archive at the Space
Telescope Science Institute. STScI is operated by the Association of
Universities for Research in Astronomy, Inc. under NASA contract NAS
5-26555. This work is based on observations carried out with the IRAM
Plateau de Bure Interferometer. IRAM is supported by INSU/CNRS
(France), MPG (Germany) and IGN (Spain).

\bibliographystyle{yahapj}

\end{document}